**Title: Versatile non-diffracting beams generator based on free lens modulation**

*Xue Yun[1], Yansheng Liang[1*], Minru He[1], Lingquan Guo[1], Xinyu Zhang[1], Tianyu Zhao[1], Ming Lei[1, 2**]*

[1]MOE Key Laboratory for Nonequilibrium Synthesis and Modulation of Condensed Matter, Shaanxi Province Key Laboratory of Quantum Information and Quantum Optoelectronic Devices, School of Physics, Xi'an Jiaotong University, Xi'an 710049, China
[2]State Key Laboratory of Electrical Insulation and Power Equipment, Xi'an Jiaotong University. Xi'an 710049, China
E-mail:
*yansheng.liang@mail.xjtu.edu.cn
**ming.lei@mail.xjtu.edu.cn

Non-diffracting beams, notable for their self-healing properties, high-localized intensity profiles over extended propagation distances, and resistance to diffraction, present significant utility across various fields. In this letter, we present a versatile and highly efficient non-diffracting beams (NDBs) generator predicated on the Fourier transformation of the focal plane field produced through the free lens modulation. We demonstrate the experimental generation of high-quality Bessel beams, polymorphic generalized NDBs, tilted NDBs, asymmetric NDBs, NDBs array and specially structured beams formed by the superposition of co-propagating beams. The versatile generalized NDBs generator is anticipated to find applications in laser processing, optical manipulation, and other fields.

## *1. Introduction*

The exploration of non-diffracting beams (NDBs) is a fascinating area in optical physics. The origin of nondiffracting modes dates back to 1987, when J. Durnin realized a special Bessel-type solution based on the cylindrical coordinate Helmholtz equation and successfully generated the Bessel beams in experiments[1, 2]. Various types of nondiffracting waves were subsequently derived by solving the wave equation in different coordinate systems, such as Mathieu beams with elliptic and hyperbolic intensity modes[3], Airy beams that transmit along curved paths[4], and transverse parabolic beams[5]. These beams exhibit the pivotal attribute of self-healing properties and maintain constant intensity profiles over extended distances, effectively circumventing traditional diffraction limits[6, 7]. Characterized by these intriguing properties, NDBs have found applications across multiple fields. For example, NDBs are effective in improving penetration into scattering media at greater distances, allowing dielectric particles or cells to be optically trapped in multiple planes simultaneously[8, 9]. Tilted Bessel beams with propagation directions of the optical axis can be used for optical delivery, laser processing, and two-photon fluorescence stereo microscopy [10-12]. In the field of biomedical imaging, Airy beams with asymmetric excitation patterns and diffraction-free properties provide higher fluorescence contrast and a larger resolution field of view compared with fast diverging Gaussian beams[13, 14].

The widespread appeal of nondiffracting beams in diverse applications has led to extensive research into their generation. The non-diffracting beam can be understood as an interference field produced by the interference of plane waves, of which the propagation vectors form a conical surface. Based on this principle, specialized optical components have been developed to realize NDBs. J. Durnin initially generated zero-order Bessel beams experimentally by using a circular slit [2]. This method is straightforward but comes with the drawback of low efficiency. Subsequently, a range of optical components such as axicons[15], meta-surfaces[16, 17], optical fibers [18], integrated silicon photonic chips [19], and enclosed cylindrical lenses [20] have also been developed as innovative techniques for generating NDBs. However, the light field structure of the NDBs produced by these elements is limited to a few modes. Various methods based on computational holography to modulate the amplitude and phase of the angular spectrum have been proposed, making the transverse intensity pattern of diffraction-free beams customizable [21-23]. Nevertheless, the complex amplitude modulation of the light field leads to significant power loss, unfavorable for applications such as optical trapping[24].

In this paper, we present a versatile NDBs generator utilizing a phase-only spatial light modulator (SLM). The generator is accomplished by applying a Fourier transform to the polymorphic focal plane fields generated via free lens modulation[25], offering high flexibility, high power usage, and straightforward experimental operation. The functionality of this generator involves producing stable, high-order, long-range Bessel beams without the necessity for specialized optical components or spiral phase plates. It also enables the generation of polymorphic generalized NDBs with transverse intensity patterns that extend beyond concentric circles to exhibit various geometric shapes. Additionally, high-quality tilted NDBs with adjustable tilt angles and other special structured light beams can also be achieved.

## 2. Methods

The first key technique of the NDBs generator is to flexibly generate various shapes of polymorphic beams on the focal plane with free lens modulation, encompassing geometrical shapes such as triangle, square, pentagonal, and oval (**Figure 1**b). Specifically, the transmission function of the digital lens can be expressed as

$$t(r, \varphi) = P(r, \varphi) e^{-i\pi(r-\rho_0(\varphi))^2/\lambda f_0} e^{im\varphi}, \tag{1}$$

which can be regarded as a combination of the phase elements from a free lens and a vortex, as illustrated in Figure 1c. Here, $P(r, \varphi)$ refers to the aperture function, and $(r, \varphi)$ denotes the polar coordinates. $\lambda$ is the wavelength of the laser, $m$ represents the topological charge, $\rho_0(\varphi)$ controls the shape of the digital free lens, and $f_0$ is the focal distance. The angular spectrum of Bessel functions forms a thin ring. Hence, a bright ring beam of tunable topological charge is required. This is achieved by using an annular free lens, as shown in Figure 1a. In this case, the predesigned radius $\rho_0(\varphi)$ is a constant, denoted as $\rho_0$.

The second key point of NDBs generator involves performing a Fourier transform on the bright polymorphic beam. This is accomplished by placing a physical lens at the focal plane of the free lens (Figure 1a). When the annular beam possesses a flat phase profile, a zero-order Bessel beam is obtained behind the physical lens. The generation of higher-order Bessel beams only requires a simple adjustment of the parameter "$m$" in Equation (1).

Figure 1d shows a schematic diagram of the experimental setup of a versatile NDBs generator. A laser beam with a wavelength of 633 nm passes through a half-wave plate (HWP) and a polarizing beam splitter (PBS), after which the beam's polarization state is converted to horizontal polarization. By rotating the fast axis direction of the HWP, the laser power is adjusted accordingly. Subsequently, the linearly polarized light is incident at a small angle (≤6 degrees) onto a phase-only SLM (HDSLM80R Plus, UPOLabs, China) with a resolution of 1920×1200, an effective area of 15.42×9.66 mm, and pixel pitch of 8 μm. Here, the parameters of the annular free lens are set as $f_0$=100 mm, $\rho_0$=0.45 mm, and $m$=1. The laser beam is modulated and focused at the focal plane of the free lens to form a thin annular beam (Figure 1e). A spatial filter is placed on the focal plane to block the zero-order beam. A physical lens (focal length $f_1$=300 mm) is placed 400 mm away from the SLM. A camera is mounted on an optical rail with a length of 110 cm to translate along the optical axis and retrieve the corresponding longitudinal intensity evolutions of the NDBs. Figure 1f shows the transverse intensity distribution of the first-order Bessel beam recorded at $z$ = 10 cm behind the physical lens and the corresponding radial intensity profile.

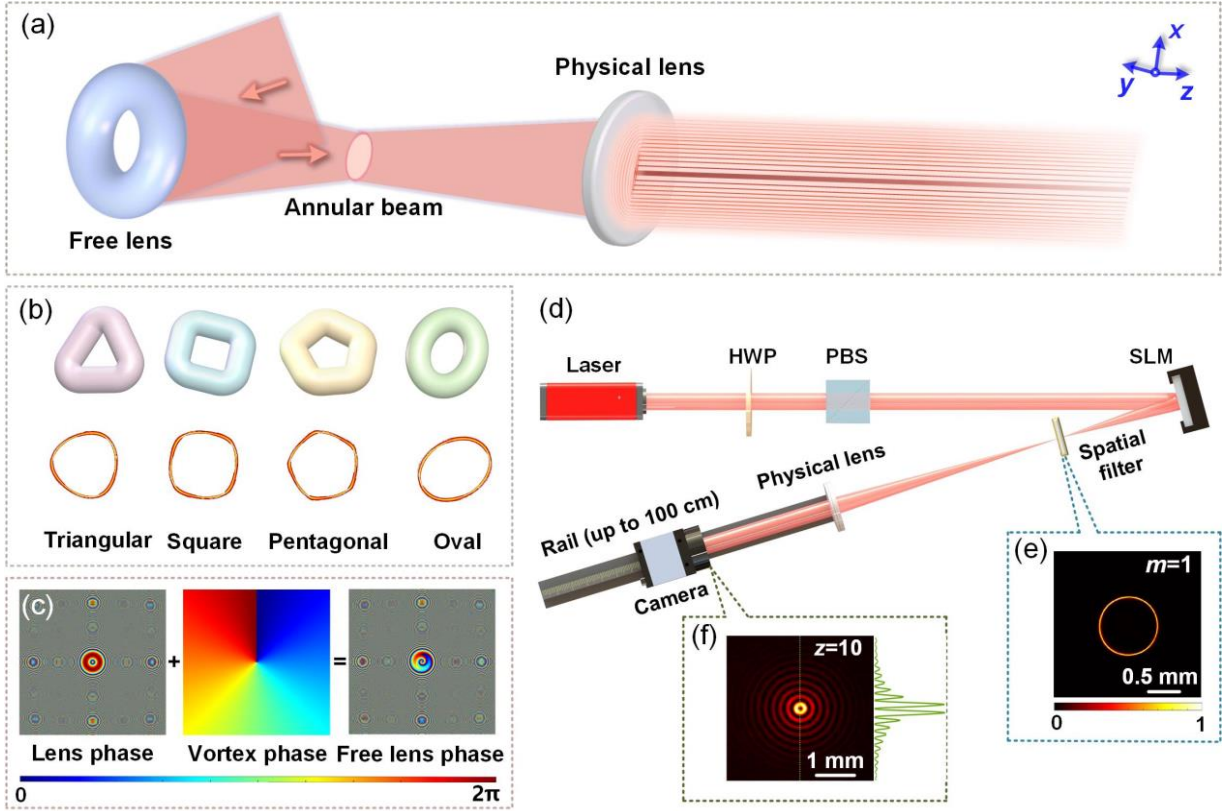

**Figure 1.** Principle of versatile NDBs generator based on free lens modulation. (a) Abridged general view of the method. (b) Polymorphic free lenses and the corresponding focal plane field profiles. (c) Superposition of the annular lens phase and the vortex phase to generate the free lens phase hologram. (d) The schematic of the experimental setup. HWP, half-wave plate; PBS, polarization beam splitter; SLM, spatial light modulator; (e) The experimental intensity pattern of an annular beam ($m$=1) at the focal plane of the free lens. (f) The transverse intensity distribution and the corresponding radial intensity profile of the experimentally produced first-order Bessel beam at $z$=10 cm.

## 3. Results and discussion
### 3.1. Non-diffracting Bessel beams

Compared with methods such as axicon and circular slit, the proposed generator does not require a trade-off between performance and flexibility. This is attributed to the formidable capabilities of the SLM and the adaptability of the free lens modulation, allowing us to create annular patterns with varying radii in experiments and achieve rapid pattern switching. As a result, we can methodically investigate and control parameters such as the non-diffracting distance and Full Width at Half Maximum (FWHM) of the produced Bessel beams. In **Figure**

**2**, we analyzed the differences between the annular beams modulated by four different free lenses and the resulting Bessel beams. Figure 2a displays the phase holograms of free lenses, with $\rho_0$ set to 0.315 mm, 0.45 mm, 0.9 mm, and 1.35 mm, respectively, and without the addition of the vortex phase. The recorded annular light fields at the focal plane of the free lenses are depicted in Figure 2b. Figure 2c displays the transverse intensity patterns of the zero-order Bessel beams at $z = 10$ cm, clearly showing an observable trend: the central peak width of the beam narrows as $\rho_0$ increases. Figure 2d portrays the trend in normalized central peak intensity for zero-order Bessel beams engendered by the four free lenses during propagation. The intensity values exhibit a gradual decrement with the extension of the propagation distance. We recorded the evolution of the Bessel beams over a normalized intensity range from 1 to 0.13. Figure 2e indicates that within this delineated propagation range, the FWHM of the central peak for the zero-order Bessel beams remains almost constant. Considering the intensity attenuation, in this paper we delineate the non-diffracting propagation distance as the interval wherein the normalized intensity value declines from 1 to 0.13. Figure 2f shows a negative correlation between $\rho_0$ and the non-diffracting range of Bessel beams. For $\rho_0$=0.315mm, 0.45mm, 0.9 mm, and 1.35 mm, the corresponding non-diffracting distances are approximately 102 cm, 95 cm, 77 cm, and 61 cm. The decreasing trend between the propagation distance and the ring size $\rho_0$ arises from the finite aperture for shaping the input beam. Consequently, a larger value of $\rho_0$ leads to a larger cone angle and a smaller interference volume, thus a shorter nondiffracting propagation distance.

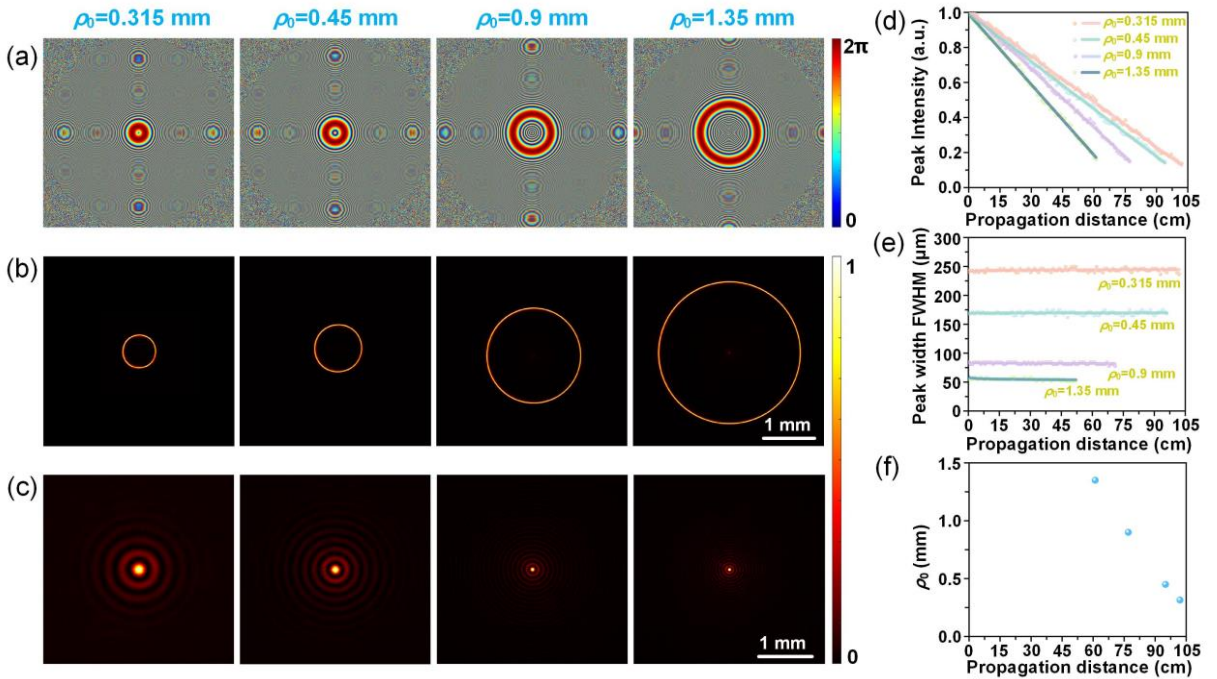

**Figure 2.** Experimentally generated zero-order Bessel beams by using annular free lenses with different $\rho_0$. (a) The free lens phase holograms with $m=0$, $\rho_0 = 0.315$ mm, 0.45 mm, 0.9 mm, and 1.35 mm, respectively. (b) The experimental intensity patterns of the annular beams at the focal plane of the free lenses. (c) The transverse intensity patterns of Bessel beams generated at $z=10$ cm. (d) Variation of the normalized peak intensity of the corresponding Bessel beams with propagation distance. (e) Relationship between the peak width FWHW of Bessel beams and the propagation distance. (f) Relationship between $\rho_0$ and the propagation distance of Bessel beams.

**Figure 3** shows the experimental results of high-quality zero-order and high-order Bessel beams produced by annular free lenses with $\rho_0$ set to 0.45 mm. Figures 3a-3c display the longitudinal intensity profiles of zero-order, first-order, and $10^{th}$-order Bessel beams propagating along the optical axis within the range from $z = 0$ cm to $z = 95$ cm from the physical lens. To better visualize the outer-lying satellite rings' transformation, we intentionally raised the laser power slightly before capturing the footage, resulting in an overexposure of the camera. It is evident that for the zero-order and first-order Bessel beams, both the central peak/ring and the outer-lying satellite rings effectively resist diffraction spreading, achieving nearly non-diffractive propagation. For the $10^{th}$-order Bessel beam, the diameters of all the rings start to slowly decrease after propagating approximately 75 cm. Figures 3d-3f present the transverse intensity patterns of the three beams at $z = 10$ cm. Figures 3g-3i show the experimental radial intensity profiles (red dashed lines) and the corresponding numerical fitted curves of the Bessel beams (blue solid lines). It can be observed that the experimentally produced Bessel beams conform to the perfect Bessel distribution. For $10^{th}$-order Bessel beam, the intensity of the outermost satellite rings decreases more rapidly, and there is a certain deviation in the positions of these rings from the theoretical results. The reason could be the modulation aberration due to the pixelation effect of SLM.

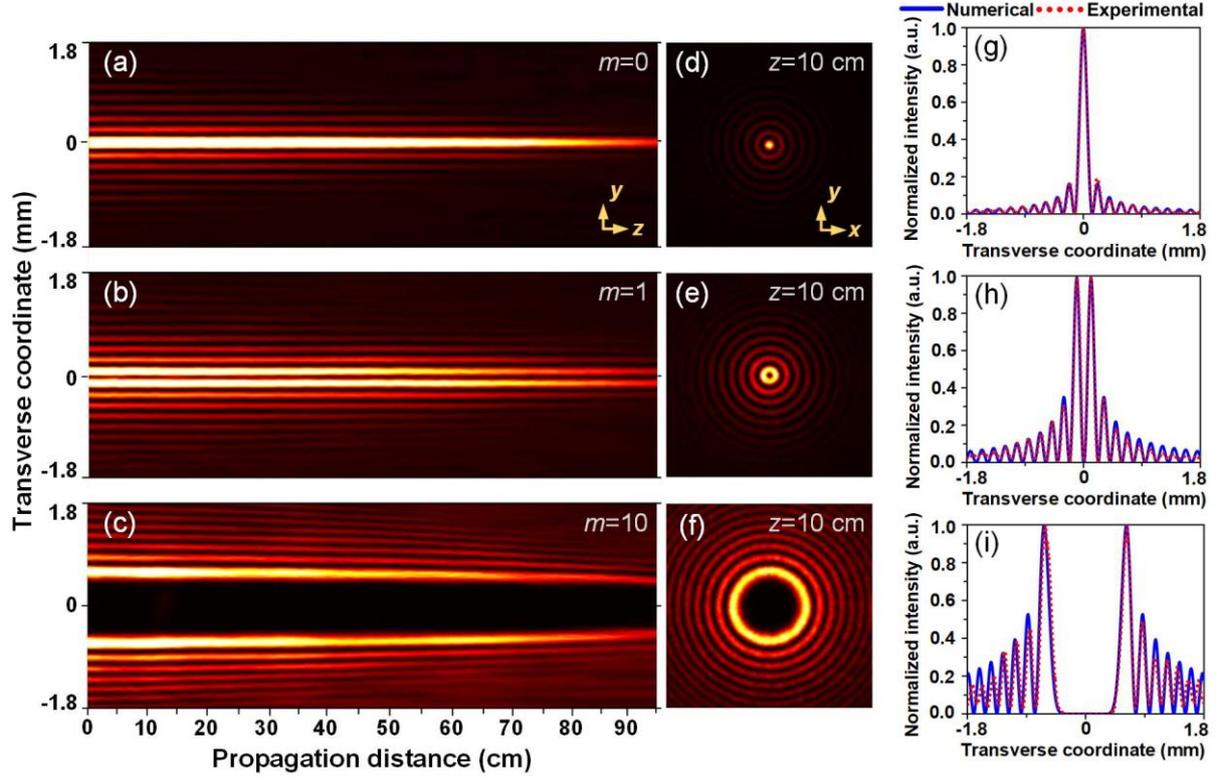

**Figure 3.** Experimentally generated zero-order, first-order, and 10$^{th}$-order Bessel beams by using free lenses with $\rho_0$=0.45 mm. (a-c) Longitudinal progression of the transverse cross-sections of Bessel beams at distances ranging from $z$=0 cm to $z$=95 cm behind the physical lens. (d-f) The transverse intensity profile of these three Bessel beams at $z$=10 cm. (g-i) The comparisons between the experimental transverse cross-sectional intensity distributions of the Bessel beams (shown as red dashed lines) and the theoretically transverse intensity distributions (depicted as blue solid lines).

### 3.2. Polymorphic generalized non-diffracting beams

An advantage of free lens modulation is evident in the flexibility to adapt the shape of the lenses, enabling the generation of a polymorphic focal plane fields. In this case, the parameter $\rho_0(\varphi)$ in Equation 1 is no longer a constant but rather follows the following relationship:

$$\rho_0(\varphi)=1-\frac{1}{p}\cos(q\varphi). \tag{2}$$

Here, $p$ determines the level of smoothness of the polygon, while $q$ determines the geometric shape of the modulated light field. In **Figure 4**a, by setting the appropriate parameters ($p$, $q$), we get the polymorphic free lenses with different shapes, such as triangular lenses ($p$=10, $q$=3) with $m$=0, 1 and 10, square lenses ($p$=15, $q$=4) with $m$=10, pentagonal lenses ($p$=20, $q$=5) with $m$=10, and oval lenses ($p$=7, $q$=2) with $m$=10. The light field intensity patterns at the focal plane of the free lenses are shown in Figure 4b. Figure 4c shows the transverse intensity patterns corresponding to the novel beams taken at different distances behind the physical lens. Figure 4d shows the three-dimensional reconstructions of the 10$^{th}$-order beams. Slices S1~S4 display the intensity patterns in some transverse planes during the propagation. The evolution of the beams is shown in Visualization 1. Zero-order and high-order beams exhibit concentric geometric shapes similar to Bessel beams and are characterized by long-range propagation. Uniquely, the intensity patterns of these beams are gradually distorted and rotated with different angles during propagation. Take the oval beam with the strongest rotational tendency as an example, at $z$=5 cm, the angle ($α_1$) between the long axis of the oval beam and the $x$ direction is about 137.6 degree, at $z$=95 cm, the angle ($α_2$) is about 77.3 degree, that is, the long axis of the oval is rotated about 60.3 degree in the clockwise direction (Figure 4c). Therefore, we refer to these novel long-distance propagation beams as polymorphic generalized NDBs. We analyzed the relationship between the topological charge value $m$ and the rotation angle $α_1$-$α_2$ of the oval generalized NDB within the propagation range from z = 5 to z = 95 cm, as shown in Figure 4e. When $m$ ranges from 5 to 15, the rotation angle decreases rapidly, and as $m$ exceeds 15, the variation of the rotation angle tends to level off. Owing to its intriguing properties, we believe that polymorphic generalized NDBs hold potential applications in imaging, high-capacity optical communication, and various other fields[26, 27].

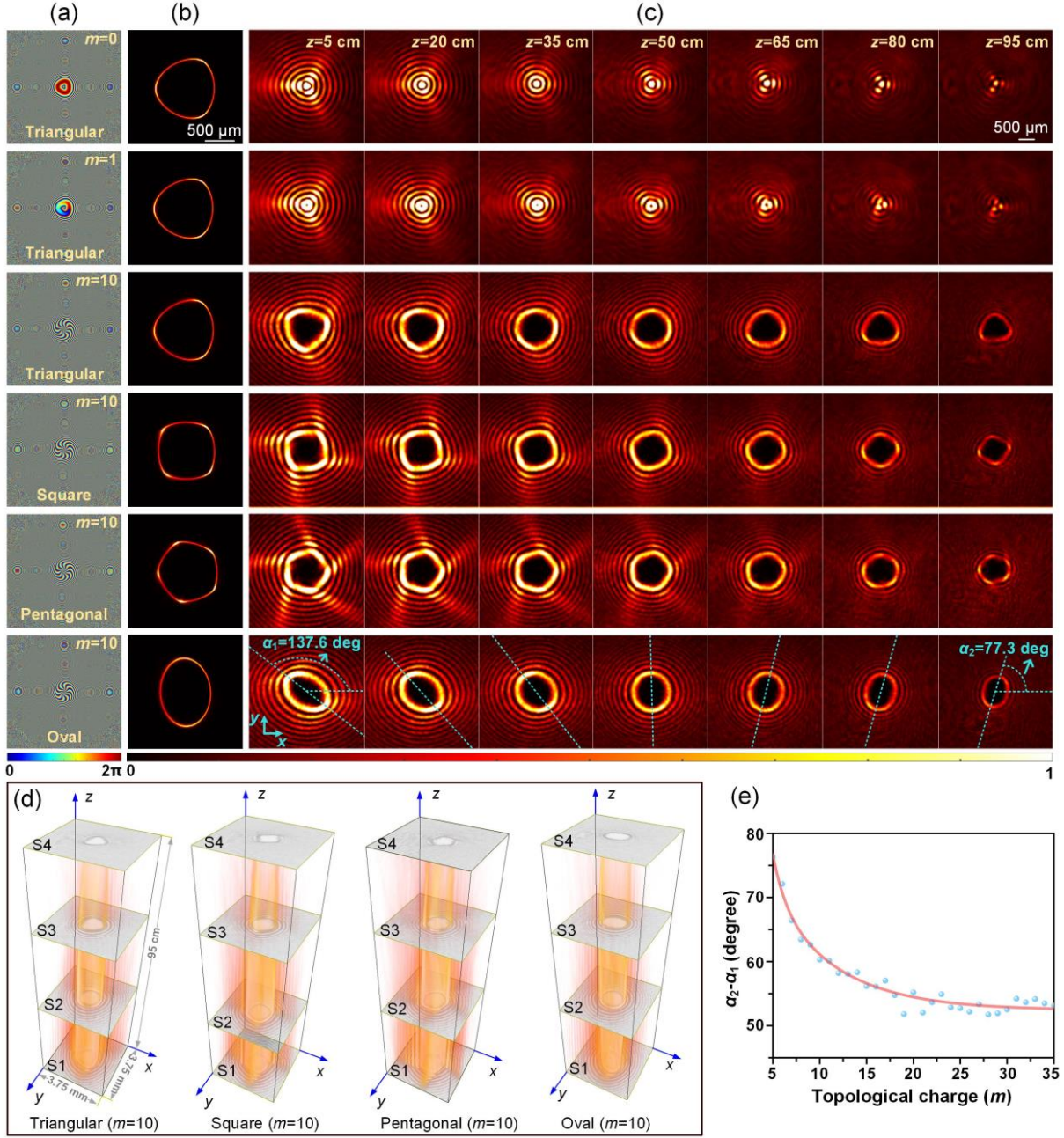

**Figure 4.** Experimentally generation of polymorphic generalized NDBs (see Visualization 1). (a) The phase holograms of free lenses. (b) The light field intensity patterns at the focal plane of the free lenses. (c) The transverse strength distribution of polymorphic generalized NDBs at different positions in the *z*-direction. (d) Three-dimensional light field reconstructions of 10[th]-order polymorphic generalized NDBs. (e) The relationship between topological charge value *m* and the rotation angle $\alpha_1$-$\alpha_2$.

### 3.3. Tilted non-diffracting beams

The significance of tilted non-diffracting beams cannot be underestimated in various fields, including optical trapping[10], biomedical photonics imaging[28, 29], laser cutting[11], and more. One technique for deviating Bessel beams from the spatial optical axis involves illuminating the axicon with incident light at an inclined angle, but precise correction of aberrations introduced by oblique incidence necessitates using a second SLM[30]. Based on the metasurface, it is possible to generate Bessel beams with any inclination angle but at the expense of flexibility[31].

As shown in **Figure 5**, the versatile NDBs generator employs the decentered free-lens technique to generate tilted NDBs with easily controlled tilt angles effortlessly. Figures 5a-5f shows the results of the experimentally generated tilted zero-order Bessel beams. Specifically, the resolution of the phase hologram is 1920×1200. We establish a coordinate system with the central pixel as the origin (the yellow dot *o*) on the hologram, as shown in Figure 5a. For the Bessel beam propagating along the optical axis, the center of the free lens is located at the origin. To generate tilted Bessel beams, altering the center position ($u_l$, $v_l$) of the free lens is required. For example, we maintain $v_l$ =0 while setting $u_l$ =180 pixels. The predesigned radius $\rho_0$ of the free lens is 0.45 and $m$=0. In this case, the intensity pattern of the annular beam on the focal plane of the free lens exhibits non-uniformity along the azimuthal direction (Figure 5b). After the Fourier transform, the tilted zero-order Bessel beam is generated behind the physical lens. We recorded the evolution of the tilted beam along the *z*-direction from 0 cm to 95 cm. The three-dimensional optical field reconstruction is depicted in Figure 5c. Slices S1 to S4 illustrate some transverse intensity distributions during propagation, displaying distinct concentric ring patterns. Figure 5d demonstrates that the FWHM of the central peak of the corresponding tilted zero-order Bessel beam exhibits remarkable stability during propagation, validating that the generated tilted beam maintains its high-quality nondiffracting characteristics. Figure 5f represents schematic diagrams of the tilted zero-order Bessel beam on the x-z plane with various lateral displacements $u_l$ of 0, 60, 120, 180, and 240 pixels. The tilt angles *θ* are approximately 0, 0.08, 0.16, 0.24, and 0.32, respectively, indicating a linear increase in the tilt angle *θ* with the parameter $u_l$, as shown in Figure 5e. Therefore, we can flexibly control the tilt angle *θ* of the Bessel beams relative to the z-axis by simply adjusting the center pixels ($u_l$, $v_l$) of the free lenses. This approach is also universally applicable to both higher-order Bessel beams and polymorphic generalized NDBs. Figures 5g and 5h show the three-dimensional volumetric reconstructions of tilted $5^{th}$-order Bessel beam with ($u_l$, $v_l$) = (-60, 60) and tilted $10^{th}$-order square generalized NDB with ($u_l$, $v_l$) = (60, -60), respectively.

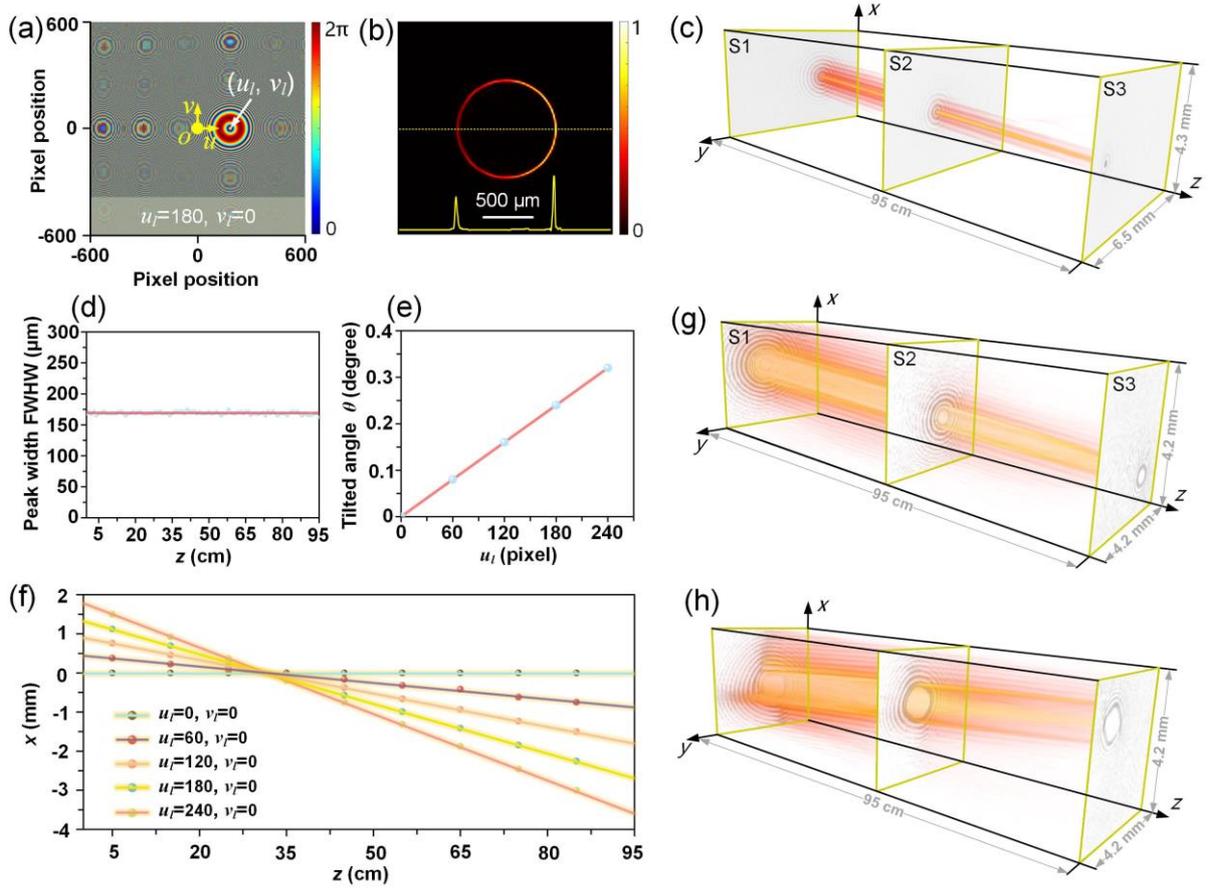

**Figure 5.** Experimentally generation of tilted NDBs. (a) The off-axis free lens phase hologram for generating a tilted zero-order Bessel beam. The center of the free lens is located at coordinates ($u_l$, $v_l$), with the yellow dot $o$ serving as the origin of the coordinate system ($u$, $v$). (b) Intensity distribution of the annular beams at the focal plane of the free lens. (c) The 3D intensity profile of the tilted zero-order Bessel beam. The slices S1-S4 show the transverse intensity distribution at various axial locations. (d) Relationship between the peak width FWHW of tilted zero-order Bessel beams and the propagation distance. (e) The tilt angle $\theta$ of the zero-order Bessel beam as a function of the off-axis displacement $u_l$ in the entrance pupil plane. (f) Schematic of the propagation of tilted zero-order Bessel beams corresponding to different $u_l$ (0, 60, 120, 180, 240 pixels, respectively) in the $y$-$z$ plane. (g-h) The volumetric reconstruction of tilted 5[th]-order Bessel beam and tilted 10[th]-order square generalized NDB. See Visualization 2 for a more visual impression.

### 3.4. Asymmetric non-diffracting beams

The asymmetric Bessel mode were theoretically proposed by V. V. Kotlyar et al.[21]. The transverse profile intensity of this beam has azimuthal variant intensity, resembling the shape of a crescent. Owing to its unique optical properties, the asymmetric Bessel beam offers potential applications in fields such as optical manipulation and microlithography. The versatile NDBs generator demonstrates a new scheme for generating asymmetric Bessel beams and extends these beams into a wide family of asymmetric polymorphic generalized NDBs with various intensity profiles, as shown in **Figure 6**. The method is based on the Fourier transform of the three-dimensional tilted focal plane field generated by the free lens. In this case, the focal length of the free lens in Equation (1) should be

$$f(\varphi)=f_0\frac{a-\sin(\varphi)}{a}, \tag{3}$$

where $a$ control the slope of the focal plane field. The tilted angle of the focal plane field decreases nonlinearly with the increase of $a$ [25]. Figures 6 (a) and (b) are the light field models of the focal plane of two free lenses with slope parameters of 2 and 4. The corresponding simulation results of the intensity and phase of first-order asymmetric Bessel beams are shown in Figures 6c-6f. The results illustrate a smooth and continuous non-uniform distribution of intensity along the central ring of beams. The dependence of the transverse intensity profiles of the asymmetric Bessel beams on the slope parameter $a$ was revealed by measuring the normalized intensity (Figure 6g). As $a$ increases from 2 to 6, the intensity of the left side of the center ring of the beams increases nonlinearly. Figures 6 (h-j) respectively show the intensity modes at different propagation distances for experimentally generated first-order asymmetric Bessel beam, 5[th]-order asymmetric triangular generalized NDB, and 10[th]-order asymmetric square generalized NDB with $a=2$. These beams each demonstrate distinct non-uniform intensity distribution features, confirming the high feasibility of the method in producing polymorphic asymmetric generalized NDBs.

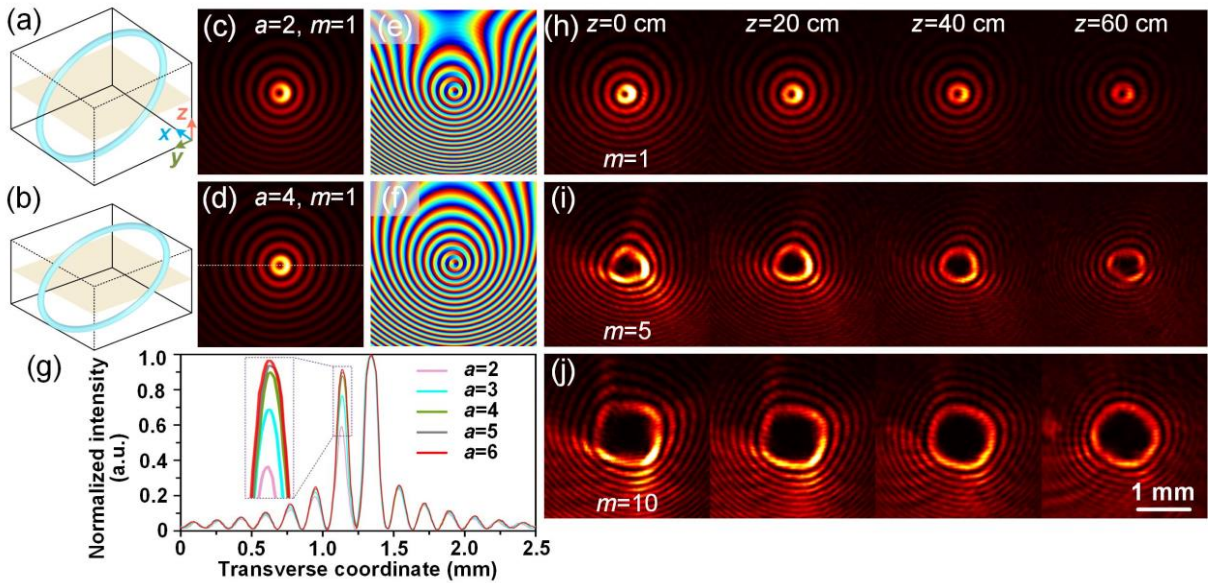

**Figure 6.** Generation of asymmetric NDBs. (a, b) Three-dimensional tilted annular ring light field models. (c, d) Simulated intensity patterns of the asymmetric first-order Bessel beams with *a* of 2 and 4, respectively. (e, f) The corresponding phase. (g) Intensity profiles of *a*=2, 3, 4, 5, and 6. (h-j) Transverse intensity profiles of the experimentally generated first-order asymmetric Bessel beam, $5^{th}$-order asymmetric triangular generalized NDB, and $10^{th}$-order asymmetric square generalized NDB at different propagation positions.

### 3.5. *Non-diffraction beams array*

Compared to the generation of a single NDB, the multi-NDBs array with capability of parallel manipulation and large capacity has attracted considerable attention and found widespread application in fields such as two-photon laser scanning stereomicroscopy and microparticle guiding [32, 33]. The versatile NDBs generator produces NDBs array by superimposing multiple light fields and extracting the phase, as shown in **Figure 7**. These light fields are generated by free lenses located at different positions of the SLM. By employing the superposition of complex fields, rather than dividing the SLM into many sub-regions to provide multiple free lens phase masks, intensity variations within the array caused by uneven illumination of the incident beam can be circumvented. Figures 7a-7c show the phase holograms used to generate the 3×3 zero-order Bessel beams array, the five-component zero-order Bessel beams array, and the three-component $5^{th}$-order polymorphic NDBs array, respectively. The central coordinates of the free lenses ($u_l$, $v_l$) have been marked in the patterns and $\rho_0$ set to 0.45 mm. The corresponding simulation results of the light fields at the focal plane of the free lenses and the transverse intensity patterns of the NDBs array are shown in Figures 7d-7f and Figures 7g-7i, respectively. By analyzing the intensity distribution curves of the three Bessel beams located on the diagonal, as marked by the yellow dashed lines in Figures 7g and 7h, it is observed that compared to the five-component Bessel beam array (Figures 7k), the 3×3 Bessel beam array (Figures 7j) exhibits relatively poorer beam quality and uniformity. This is due to mutual interference caused by the proximity of neighboring spots. Therefore, it is feasible to decrease the number of spots or increase the distance of the two neighboring beams by adjusting the relative central coordinates of the free lenses to increase uniformity. Figure 7l shows the three-dimensional volumetric reconstruction of the experimentally generated 2×2 zero-order Bessel beams array. It is noteworthy that although this method offers the advantages of high cost-efficiency and high flexibility, it is not suitable for applications that require the Bessel beams in the array to be parallel to each other along the direction of propagation.

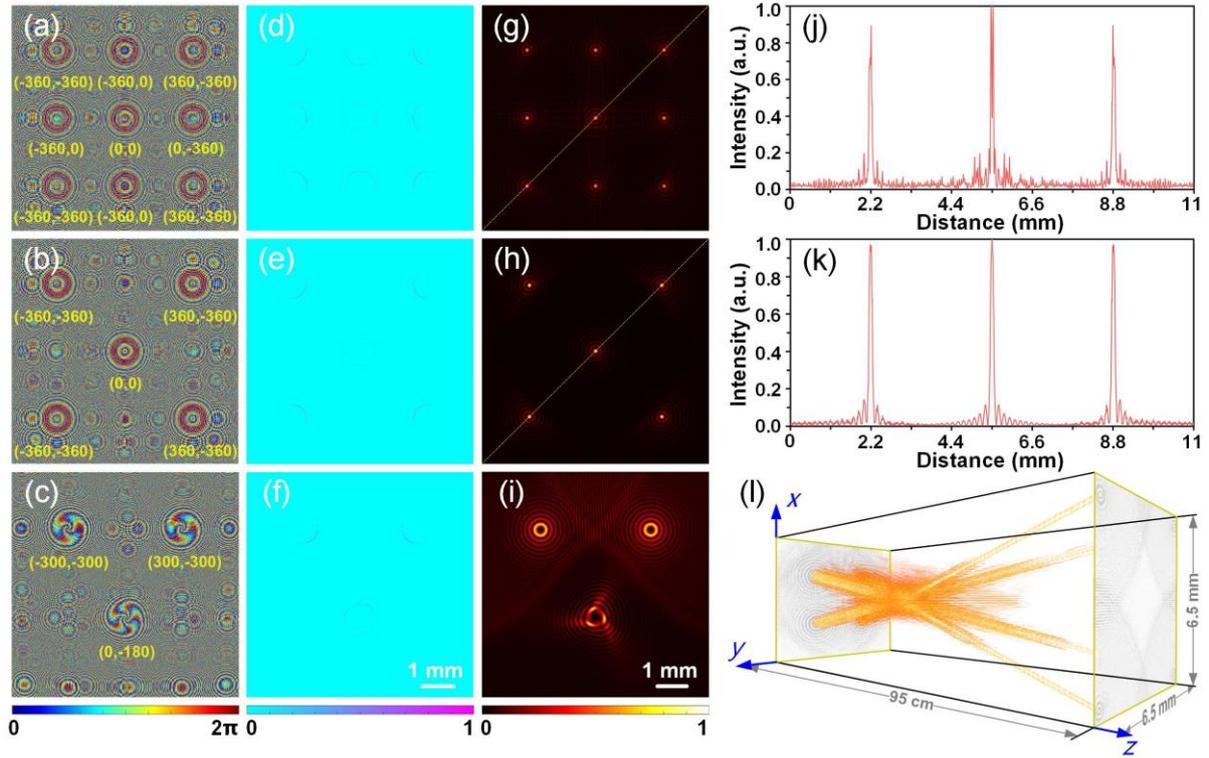

**Figure 7.** Generation of NDBs array. (a-c) The phase patterns for generating 3×3 zero-order Bessel beams array, five-component zero-order Bessel beams array, and three-component 5$^{th}$-order polymorphic generalized NDBs array. The central coordinates ($u_l$, $v_l$) of the free lenses are indicated in the patterns. (d-f) The corresponding simulation results of the field distribution at the focal plane of the free lenses. (g-i) The simulation results of the transverse intensity pattern of the arrays at $z$=0. (j-k) Intensity distributions of the three Bessel beams marked by yellow dashed lines in (g-h). (l) Experimental results of volumetric reconstruction of 2×2 zero-order Bessel beams array (Visualization 3).

### 3.6. Superpositions of Bessel beams

The superposition of co-propagating Bessel beams can generate some distinctive structured light beams, such as circular spots array[34], helical beams [35] and optical conveyor beams [36]. The circular spots array consists of controlled number of diffraction-free petal-like spots arranged on a circular intensity pattern of radius associated with $m^{th}$-order Bessel function. Helical beams manifest a continuous helical trajectory throughout their propagation and are a subset of radially self-accelerating beams. Optical conveyor beams display a sequence of focal spots aligned along the propagation vector that act as efficient optical traps for particles. Experimentally, the superposition of Bessel beams can be achieved by employing two concentric annular free lenses, as depicted in the first row of **Figure 8**. This arrangement yields interfering ring or concentric double-ring light patterns on the focal plane (the second row of Figure 8). The fourth row shows the volumetric reconstruction of the experimentally generated circular spot array, helical beams and conveyor beams. The transverse intensity patterns measured at a distance of 10 cm are shown in the third row. Figures 8a and 8b show the circular spot arrays generated by two spatially overlapping higher-order Bessel beams ($\rho_0$ set to 0.45 mm) with topological charges of the same magnitude but opposite signs. The number of diffraction-free spots and the diameter of the dot array are related to the magnitude of the topological charges. When the absolute value of $m$ is 3, the number of spots in the array is 6 (Figure 8a) and it increases to 10 when the absolute value is 5 (Figure 8b). Figure 8c shows the two free lenses with $\rho_0$ set to 0.45 mm and 0.9 mm and corresponding phase orders $m$ of -1 and +2. The generated helical beam is featured with threefold rotational symmetry. By modulating the $m$ of the outer ring to +3, a four-component helical beam can be easily realized (Figure 8d). Additionally, the optical conveyor beam is generated based on two free lenses with flat phase free lenses (Figure 8e). Here, $\rho_0$ is set at 0.45 mm and 1.35 mm, respectively, and within a range of 53 cm, 8 high-intensity focus-like spots are measured. The spacing between adjacent spots along the propagation direction progressively increases while the spots gradually decrease in size. In Figure 8f, we reset the $\rho_0$ corresponding to the outer ring at 1.2 mm and fixed the center coordinates ($u_l$, $v_l$) of free lens at (-120, 0), resulting in a tilted optical conveyor beam with focal spots of about seven. The tilt angle $\theta$ relative to the optical axis is about 0.16 degrees.

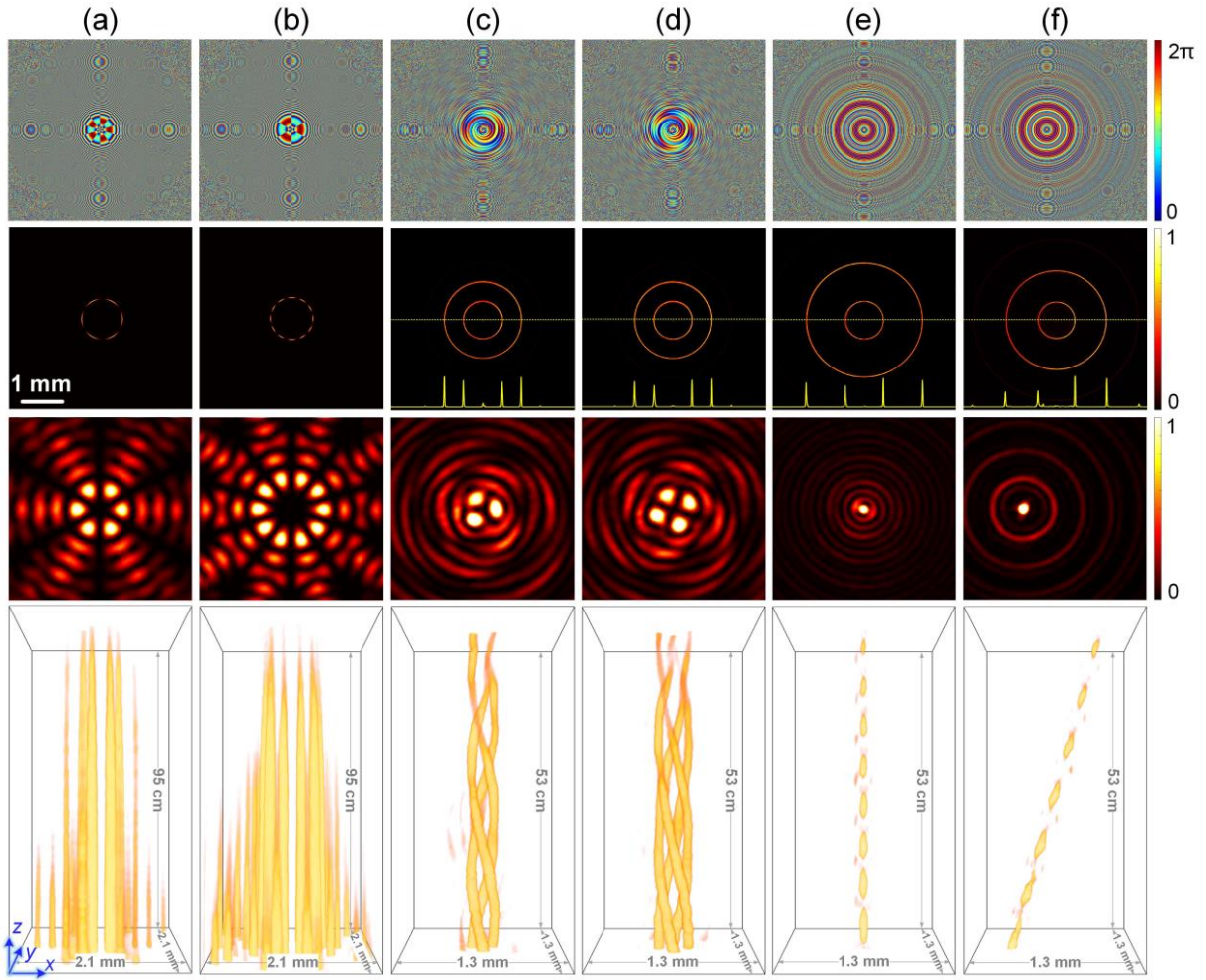

**Figure 8.** Experimentally generation of circular six-spot array (a), circular ten-spot array (b), three-component helicon beam (c), four-component helicon beam (d), optical conveyor beam (e), and tilted optical conveyor beam (f). The first row shows the phase holograms configured by two concentric free lenses. The second row depicts the light fields generated at the focal plane of the free lenses. The transverse intensity patterns measured at a distance of $z$=10 cm are shown in the third row. The fourth row presents the volumetric reconstructions of the beams. See visualization 4 for a more visual impression.

### 3.7. Self-healing

A distinctive optical property of NDBs is their remarkable self-healing capability[37]. Even when a portion of the beam is affected by scattering or diffraction, the central region of the beam can autonomously regenerate, preserving its original optical characteristics. This self-healing feature provides distinctive advantages for NDBs in various applications. To assess the self-healing capability of the generated Bessel beams and polymorphic twisted generalized NDBs, we placed a thin iron wire with a diameter of approximately 100 μm at a position $z = 3$ cm behind the physical lens to obstruct the beams. **Figure 9** illustrates the self-healing process for zero-order, first-order, and 5th-order Bessel beams (Figure 9a) and triangular generalized NDBs (Figure 9b). The beams are significantly perturbed and disrupted at the position z = 5 cm. However, after propagating for about 20 cm, these beams have largely restored their original properties, demonstrating the robust self-healing capability of the NDBs generated by the generator.

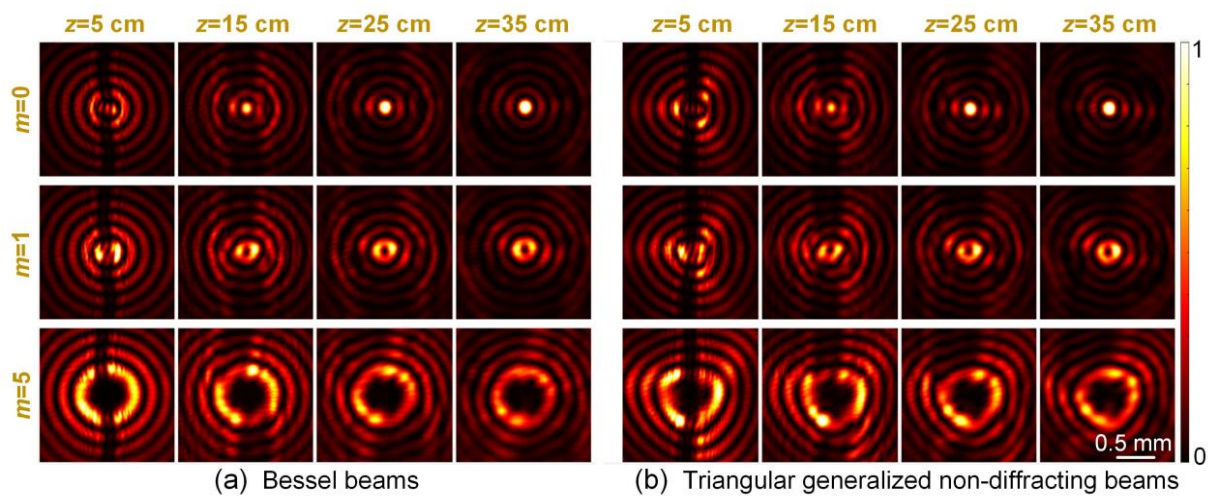

**Figure 9.** The self-healing capability of the Bessel beams (a) and the triangular generalized NDBs (b) with topological charges $m$ of 0, 1, and 5. The beam obstacle is placed at $z$=3 cm. The beams heal after $z$=25 cm.

## 4. Conclusion

In summary, a versatile, non-diffracting beams generator is proposed in this paper based on free lens modulation. The method employs a phase-only SLM without the requirement for additional specialized optical components, thereby providing enhanced flexibility and simplicity. High-quality Bessel beams with negligible transverse evolution over extended propagation distances have been experimentally produced. Furthermore, we extend the function of generator to realize polymorphic generalized NDBs, tilted NDBs, asymmetric NDBs and other special structured light beams by precisely varying the basic parameters of the free lens. Due to the significant advantages of great flexibility and high-power usage, predictably, the demonstrated generator holds potential applications in fields such as optical trapping, high-capacity optical communication, laser processing, and other fields.

**Supporting Information**

Supporting Information is available from the Wiley Online Library or from the author.

## Acknowledgments

This work was supported in part by the National Key Research and Development Program of China (2022YFF0712500); Natural Science Foundation of China (NSFC) (Nos. 62135003, 61905189, 62205267); The Innovation Capability Support Program of Shaanxi (No. 2021TD-57); Natural Science Basic Research Program of Shaanxi (Nos. 2020JQ-072, 2022JZ-34); and NIH grant (GM144414) to P.R. Bianco.

Received: ((will be filled in by the editorial staff))
Revised: ((will be filled in by the editorial staff))
Published online: ((will be filled in by the editorial staff))